\documentclass[twocolumn]{revtex4}

 \usepackage{amsmath}
  \usepackage{paralist}
  \usepackage{graphics} 
  \usepackage{epsfig} 
 \usepackage[colorlinks=true]{hyperref}

\hypersetup{urlcolor=blue, citecolor=red}

\usepackage{amssymb}
\usepackage{mathrsfs}
\usepackage{bm}
\usepackage{rotating}
\usepackage{cancel}
\usepackage{wasysym}
\usepackage[normalem]{ulem}
\usepackage{multirow}

\begin{document}

\preprint{APS/123-QED}

\title{On the moment system 
and a flexible Prandtl number}

\author{Ra\'ul Machado}
 \email{raul\_machado@hotmail.com} 
\affiliation{%
Faculty of Engineering and the Environment, University of Southampton, Southampton, SO17 1BJ, United Kingdom}%


\begin{abstract}
The Maxwell-Boltzmann moment system can be seen as a particular case of a mathematically more general moment \textit{system} proposed by Machado in Ref. \cite{MachadoPaper3Moments1D}. 
This last moment system, 
whose integral generating form and a suggested continuous distribution are presented here, 
is used in this work to theoretically show (one of) its usefulness: A flexible Prandtl number can be obtained in both the Boltzmann equation and in the lattice Boltzmann equation with a conventional single relaxation time Bhatnagar-Gross-Krook collision model.  
\end{abstract}


\maketitle

\section{Introduction}\label{section:IntroductionHigherOrder}

The Boltzmann equation (BE) \eqref{eq:BoltzmannEq} is used to describe the behavior of rarefied gases  
\begin{eqnarray} \label{eq:BoltzmannEq}
\partial_t f + c_{\alpha} \partial_{\alpha} f = \frac{1}{\tau^{*}}(f^{\textrm{eq}}-f),
\end{eqnarray}
where the rhs is the nonlinear Bhatnagar-Gross-Krook (BGK) \cite{BhatnagarGrossKrook} 
collision model.
The main effect of the collision model is to bring the velocity distribution function closer to the local equilibrium state. 
$f$ represents the probability of finding a particle with velocity $\boldsymbol{c}$ at a certain position $\boldsymbol{x}$ and at certain time $t$. $f^{\textrm{eq}} \equiv f^{\textrm{eq}}(\rho, \boldsymbol{u}, \theta, \boldsymbol{c})$ is the local equilibrium distribution function (EDF), with local flow velocity  $\boldsymbol{u}$ and density $\rho$. $\theta=R T$, i.e. temperature in energy units, where $R$ is the specific gas constant and $T$ is the local temperature. Hence, the local equilibrium $f^{\textrm{eq}}$ is determined by the local conservative flow variables, i.e. density, the momentum and the energy. In the BE, $f^{\textrm{eq}}$ is usually the Maxwell-Boltzmann (MB) 
distribution function (more about this below).  
$\tau^{*}$ is the single relaxation time, which represents the relaxation process from the nonequilibrium state towards to a local equilibrium $f^{\textrm{eq}}$, and it is related to the viscosity of the fluid. 
However, based on an approach to obtain the macroscopic relations, e.g. method of moments on the Boltzmann equation \eqref{eq:BoltzmannEq} and using the MB moments, 
the result are macroscopic relations 
with the dimensionless Prandtl (Pr) number equal one. This, in contradiction to the physical Pr $=2/3$ for monatomic gases, which is important in the study of heat transport phenomena.  

Several strategies are adopted to deal with that limitation, in order to get a correct Pr number. 
The ellipsoidal statistical (ES-BGK) approach is proposed in \cite{Holway63}, \cite{Holway66}, and the entropy condition (H-theorem) is later proven in \cite{AndriesESM2000}.  
The ES-BGK model contains a free parameter, which is useful to obtain the desired Pr number. 
This is achieved by only changing the stress tensor $\sigma$ 
via a flexible $b$ term. 
Another model is proposed by Shakhov in \cite{Shakhov68} (Shakhov-BGK), which only changes the 
 heat flux $q$, and the H-theorem is 
proven near local equilibrium state \cite{ZhengStruchtrup2005}. A construction, combining the aforementioned  both models, is also found in the literature, c.f. \cite{XuHuang2010}, as a sum of both ES-BGK model  and Shakhov-BGK model post-collision terms. This, in order to change both the stress tensor and heat flux simultaneously while maintaining  a free parameter, useful to get the desired Pr number. Numerical comparisons between ES-BGK and Shakhov-BGK models are found in the literature, c.f. \cite{MieussensStruchtrup2004}, \cite{MengWuReeseZhang2013}. 
In \cite{GLiu90}, a pure theoretical model is proposed by Liu, where both the stress tensor and the heat flux are changed by modifying the collision term in order to obtain an accepted Pr number. 

However, unlike the present approach, those aforementioned constructions are:  
$i)$ designed, just having in mind, the macroscopic conservative relations, 
$ii)$ not necessarily as a result of the use of a general mathematical framework. Thus, the theoretical potential to go beyond the analyzed scales  can be compromised.   
While the mean collision frequency in the ES-BGK, Shakhov-BGK and the Liu models are independent of the microscopic velocity, there are other models that use velocity dependent collision frequencies c.f. \cite{BouchutPerthame1993}, \cite{Struchtrup1997},  \cite{ZhengPhD2004}, \cite{MieussensStruchtrup2004}. 

Despite that some of the aforementioned construction have not been developed recently, there are many unanswered questions regarding their qualities, characteristics, performance, etc. 
Because of another view is presented in this work for the first time, some theoretical outcomes are outlined, and a complete treaty about  it, cannot be claimed here either. Therefore, further theoretical work 
and numerical assessments for different flows (e.g. Poiseuille, Couette) deserve separate works and are presented by the author  elsewhere. 

The use of multiple relaxation time (MRT) is also another adopted strategy to obtain a realistic Pr number. For instance, in the lattice Boltzmann (LB) method (\cite{McNamaraZanetti}, \cite{HigueraJimenez1989}, \cite{HigueraSucciBenzi1989}, \cite{Koelman1991}, \cite{ChenChenMatthaeus1992}, \cite{QianHumiereLallemand1992}, \cite{SucciBook},\cite{HanelGermanBook2004}, \cite{GuoShuBook}), 
the MRT is very popular, c.f.  \cite{Humiere1992}, \cite{LallemandLuo2003}, \cite{ShanChen2007}, \cite{PhilippiHegeleSurmasSiebertSantos2007},  
\cite{ZhengShiGuo2008}, \cite{ChenXuZhangLiSucci}, to mention few. 
 For those unacquainted with the LB method, the continuous Boltzmann equation can be particularly \textit{discretized} in both time and phase space \cite{HeLuoDerivationLB}, leading to the LBGK equation 
\begin{eqnarray} \label{eq:LB}
f_i(\boldsymbol{x}+\boldsymbol{c}_i \delta t,t+\delta t) = f_i(\boldsymbol{x},t) + \frac{1}{\tau} (f_i^{\textrm{eq}}-f_i), 
\end{eqnarray}
where $i=1, \dots, n_q$ and $n_q$ is the number of discrete lattice velocity vectors. In the LB method, the equilibrium $f^{\textrm{eq}} \equiv f^{\textrm{eq}}(\rho, \boldsymbol{u}, \boldsymbol{c}, \theta)$ is modeled, and also local. $\tau$ is the relaxation time,  non-dimensionalized with $\delta t$.  Within the LB method context, the terms $\boldsymbol{c}$, $\boldsymbol{u}$, $t$, $\boldsymbol{x}$ and $\theta$ are in lattice units (LU), c.f. \cite{SucciBook}. 
$f_i$ 
 represents the probability of finding a particle with velocity $\boldsymbol{c}_i$ at position $\boldsymbol{x}$ and time $t$ in LU.  
 Usually, the chosen denotation in the LB constructions follows the one in \cite{QianHumiereLallemand1992}, D$d$Q$(2 z +1)^d$, but here 
 the one-dimensional $z$ term (as defined in \cite{MachadoPaper3Moments1D}) is used. $d$ is the dimension in this work.  
 It should be pointed out that the use of more than one relaxation time has also a positive effect on the stability of the LB method, when compared to a single relaxation time LB method. Another strategy (in the LB method) is the use of the double distribution function approach \cite{HeChenDoolenJCP1998}  (as the opposite to single distribution approach used in this work), which includes two relaxation times (TRT). 
The TRT strategy is also followed in \cite{Shan1997}. 
A modified BGK collision model is the chosen strategy in \cite{GanXuZhangLi2011}. 
 
 Because of the LB method is a discretization of the Boltzmann equation then, it should not be a surprise that the aforecited strategies implemented within the Boltzmann equation context can also be used on the LB equation, c.f. \cite{MengZhangHadjiconstantinouRadtkeShan2013}. Another strategy, based on derivations involving the Fokker-Planck equation, is also found in \cite{GorjiTorrilhonJennyFP2011}, to obtain the correct Pr number of 2/3 for monatomic gases. However, the present work is focused on the BGK-BE \eqref{eq:BoltzmannEq} and LBGK equation \eqref{eq:LB}, to \textit{theoretically} show that they are enough to obtain a Pr number at will. 

The rest of this work is organized as follows: Section \ref{section:MomentSystem} deals with the moment system. 
The derivation of the Prandtl number is carried out in section \ref{section:Prandtl} and further discussions are found in section \ref{section:Discussions}. 

\section{On the moment system}\label{section:MomentSystem}

One can argue that all those aforementioned approaches have been efforts to circumvent the shortcomings of the MB moments. In this work the adopted strategy is based on the results obtained by Machado in \cite{MachadoPaper3Moments1D}, where 
a new hydrodynamic moment \textit{system} is proposed 
(c.f. table  VI, Eqs. 50 and 51 therein). Although a new high-order LB construction is used to derivate the hydrodynamic moment system 
in \cite{MachadoPaper3Moments1D}, the results are not necessarily  limited to such discrete constructions. 
Based on the work in \cite{MachadoPaper3Moments1D}, 
the moments can \textit{also} be obtained from the following integral relation: 
\begin{eqnarray} \label{eq:GeneralizedMoment}
\int \Bigg(\underbrace{\frac{\rho}{(2 \pi \theta \mathcal{F}_{n})^{d/2}} \textrm{\Large{\textit{e}}} \Big(-\frac{(\boldsymbol{c}-\boldsymbol{u})^2}{2 \theta \mathcal{F}_{n} }\Big)}_{I} \Bigg) \boldsymbol{c}^n d \boldsymbol{c}. \ \  
\end{eqnarray} 
The containing 
terms $\mathcal{F}_{n}^m$
(defined in Eq. (51) in \cite{MachadoPaper3Moments1D}) are repeated in Eq. \eqref{eq:FunctionWithMod}: 
\begin{eqnarray} \label{eq:FunctionWithMod}
\mathcal{F}_{n}^{m_{\textrm{Max}}} = \prod_{m=0}^{m_{\textrm{Max}}-1} \frac{n-2m-\textrm{mod}(n-2m+1,2)+2 \mu}{n-2m-\textrm{mod}(n-2m+1,2)}. \
\end{eqnarray}
The direct insertion of $\mathcal{F}_{n}^m$ for $m=1$, $\mu =0$, and $\mu \neq 0$ with $n= 2, 3$ into the $I$-term in Eq. \eqref{eq:GeneralizedMoment} gives a continuous EDF with existing variance equal to $\theta \mathcal{F}_{n}$. Examples are depicted in Fig.  \ref{fig:ContinuousMachadoDistribution}. This suggests that instead of a MB distribution (i.e. $\mathcal{F}_{n}=1$ when $\mu=0$), the obtained EDF can be used to define a continuous EDF in the BE. The integral \eqref{eq:GeneralizedMoment} can be evaluated thereafter to get the first fourth moments (counting from zero). In order to generate moments beyond the third-order, other special functions are needed in the $I$-term (c.f. \cite{MachadoPaper3Moments1D}), whose theoretical analysis is out of the scope of this work. Nevertheless, the evaluation of Eq. \eqref{eq:GeneralizedMoment} and the 
\textit{subsequent} substitution of the terms $\mathcal{F}_{n}^m$ can be used as a practical integral-based generating procedure. This, in order to match the new moments, 
and to obtain the equations concerning the conservation laws. The first five convective moments can be written in the following form: 
\begin{widetext}
\begin{subequations} \label{eq:AccuracyD1Q51D}
\begin{eqnarray} \label{eq:AccuracyD1Q51D-0}
\int f^{\textrm{eq}}(\rho, \boldsymbol{u}, \boldsymbol{c}, \theta)  d \boldsymbol{c} &=& \rho,  \\ \label{eq:AccuracyD1Q51D-1}
j_{\alpha} = \int f^{\textrm{eq}}(\rho, \boldsymbol{u}, \boldsymbol{c}, \theta)  c_{\alpha} d \boldsymbol{c} &=& \rho u_{\alpha},  \\  \label{eq:AccuracyD1Q51D-2}
P_{\alpha, \beta} = \int f^{\textrm{eq}}(\rho, \boldsymbol{u}, \boldsymbol{c}, \theta)  c_{\alpha} c_{\beta} d \boldsymbol{c} &=& \rho \Big((2 \mu +1) \delta_{\alpha, \beta} \theta + u_{\alpha} u_{\beta}\Big),  \\ \label{eq:AccuracyD1Q51D-3}
Q_{\alpha, \beta, \Omega} = \int f^{\textrm{eq}}(\rho, \boldsymbol{u}, \boldsymbol{c}, \theta) c_{\alpha} c_{\beta} c_{\Omega} d \boldsymbol{c} &=& \rho  \Bigg[ \Bigg( \Big(\frac{2 \mu}{1  + \delta_{\alpha, \Omega} + \delta_{\alpha, \beta}} + 1 \Big) u_{\alpha} \delta_{\beta, \Omega} \\ \nonumber
&+& \Big(\frac{2 \mu}{\delta_{\beta, \Omega}  + 1 + \delta_{\alpha, \beta}} +1 \Big) u_{\beta} \delta_{\alpha, \Omega} \\ \nonumber
&+& \Big(\frac{2 \mu}{\delta_{\beta, \Omega}  + \delta_{\alpha, \Omega} + 1} + 1 \Big) u_{\Omega} \delta_{\alpha, \beta}\Bigg) \theta  +  u_{\alpha} u_{\beta} u_{\beta} \Bigg],  \\ \label{eq:AccuracyD1Q51D-4}
R_{\alpha, \beta, \Omega, \Omega} = \int  f^{\textrm{eq}}(\rho, \boldsymbol{u}, \boldsymbol{c}, \theta) c_{\alpha} c_{\beta} c_{\Omega}^2  d \boldsymbol{c} &=& \rho \Big[(2 \mu + 1)(2 \mu  + \delta_{\alpha, \beta} + \delta_{\alpha, \Omega} + \delta_{\Omega, \beta}) \theta^2 \delta_{\alpha, \beta} \\ \nonumber
&+& (2 \mu + \delta_{\alpha, \beta} + \delta_{\alpha, \Omega} + \delta_{\beta, \Omega}) \theta u_{\alpha} u_{\beta} \delta_{\alpha, \beta} \\ \nonumber
&+& (2 \mu + 2\delta_{\beta, \Omega} + 1) \theta 
(u_{\alpha}  \delta_{\beta, \Omega} + u_{\Omega} (\delta_{\alpha, \beta}-\delta_{\alpha, \Omega})) 
u_{\Omega}  \\ \nonumber
&+&  u_{\alpha} \big((2 \mu +1) \theta (1-\delta_{\alpha, \beta})(1-\delta_{\beta, \Omega}) \\ \nonumber 
&+& u_{\beta}^2 \delta_{\beta,\Omega} + u_{\Omega}^2 (1-\delta_{\beta,\Omega})\big) u_{\beta} \Big]. 
\end{eqnarray}
\end{subequations}
\end{widetext}
$\delta_{\alpha, \beta}$ is the Kronecker delta, which equals to one when $\alpha = \beta$, otherwise is zero. 
$u_{\alpha}$ is the flow velocity on $\alpha$. The terms $\alpha$, $\beta$ and $\Omega$ are the three axes coordinate.
The above integral representations 
are over the velocity space and are written in \textit{condense notation}, so that for instance, in a 
two-dimensional case 
$Q_{\alpha, \beta, \beta} = \int f^{\textrm{eq}} (\boldsymbol{x}, \boldsymbol{c}, t) \boldsymbol{c}^3 d \boldsymbol{c}=\int f^{\textrm{eq}} (\boldsymbol{x}, \boldsymbol{c}, t) c_{\alpha} c_{\beta}^2 d \boldsymbol{c}$ 
means 
\begin{subequations} 
\begin{eqnarray} \nonumber
Q_{\alpha, \beta, \beta} &=& 
\rho \int_{-\infty}^{\infty} \frac{1}{(2 \pi \theta \mathcal{F}_{1})^{1/2}}\textrm{\Large{\textit{e}}}\Big( \frac{(c_{\alpha}-u_{\alpha})^2}{2 \mathcal{F}_{1} \theta}\Big)  c_{\alpha} d c_{\alpha} \\ \nonumber
&\times& \int_{-\infty}^{\infty} \frac{1}{(2 \pi \theta \mathcal{F}_{2})^{1/2}} 
\textrm{\Large{\textit{e}}}\Big( \frac{(c_{\beta}-u_{\beta})^2}{2 \mathcal{F}_{2}} \Big) c_{\beta}^2  d c_{\beta}.
\end{eqnarray}
\end{subequations}

When $\mu=0$, the last three moments  \eqref{eq:AccuracyD1Q51D} are reduced to, e.g.:  
\begin{subequations}
\begin{eqnarray} \nonumber
P_{\alpha, \alpha} &=& 
\rho ( \theta + u_{\alpha}^2),  \qquad \  \ P_{\alpha, \beta} = \rho u_{\alpha} u_{\beta}, \\ \nonumber
Q_{\alpha, \alpha, \alpha} &=& 
\rho  (3 \theta u_{\alpha} +  u_{\alpha}^3), \ 
Q_{\alpha, \beta, \beta} = 
\rho  ( \theta u_{\alpha} +  u_{\alpha} u_{\beta}^2), \\ \nonumber
Q_{\alpha, \beta, \Omega} &=& 
\rho  (u_{\alpha} u_{\beta} u_{\Omega} ), \  \
R_{\alpha, \alpha, \alpha, \alpha} = 
\rho (3 \theta^2 + 6 u_{\alpha}^2 \theta + u_{\alpha}^4), \\ \nonumber
R_{\alpha, \beta, \beta, \beta} &=& 
\rho (3 u_{\beta} \theta + u_{\beta}^3) u_{\alpha},  \ 
R_{\alpha, \beta, \Omega, \Omega} = 
\rho u_{\alpha} u_{\beta} (u_{\Omega}^2 + \theta), \\ \nonumber
R_{\alpha, \alpha, \beta, \beta} &=& 
\rho (u_{\alpha}^2 + \theta)(u_{\beta}^2 + \theta) ,  
\end{eqnarray}
\end{subequations}
which are (some of) their equivalent MB hydrodynamic moments. Recall that the above first five hydrodynamic convective moments,   seen in Eqs. \eqref{eq:AccuracyD1Q51D}, represent the density ($\rho$), momentum density ($j$), pressure tensor ($P$), energy flux ($Q$), and the rate of change of the energy flux ($R$) respectively. 
Similarly, the central moments can be obtained, using the relations \eqref{eq:AccuracyD1Q51D}, e.g.: 
\begin{subequations} \label{eq:AccuracyD2QnqCentral}
\begin{eqnarray} \label{eq:AccuracyD2QnqCentral-0} 
\int f^{\textrm{eq}}
  d \boldsymbol{c} = \rho, && \\ \label{eq:AccuracyD2QnqCentral-1} 
\int f^{\textrm{eq}}
  (c_{\alpha}-u_{\alpha})  d \boldsymbol{c}  = 0, && \\ \label{eq:AccuracyD2QnqCentral-2}
\int f^{\textrm{eq}}
  (c_{\alpha}-u_{\alpha}) (c_{\beta}-u_{\beta})  d \boldsymbol{c} 
= \rho (2 \mu +1) \delta_{\alpha, \beta} \theta, && \ \  \\ \label{eq:AccuracyD2QnqCentral-3}
\int f^{\textrm{eq}}
  (c_{\alpha}-u_{\alpha}) (c_{\beta}-u_{\beta}) (c_{\Omega}-u_{\Omega}) d \boldsymbol{c} \\ \nonumber
= -4 \rho \theta u_{\alpha} \mu \delta_{\alpha, \beta} \delta_{\beta, \Omega}, && 
\end{eqnarray}
\end{subequations}
where Eq. \eqref{eq:AccuracyD2QnqCentral-3} is not always zero as in the MB moments. 
Note that in the new moment system (c.f. Eqs. \eqref{eq:GeneralizedMoment}, \eqref{eq:AccuracyD1Q51D}, \eqref{eq:AccuracyD2QnqCentral} and (50)-(51) in \cite{MachadoPaper3Moments1D}) \textit{all} the hydrodynamic moments from the second-order 
and onward are modified when compared to the MB system.
The shown rhs outputs in Eqs. \eqref{eq:AccuracyD1Q51D} and \eqref{eq:AccuracyD2QnqCentral} has been algebraically verified with the D$d$Q$(2z+1)^d$ lattice set using the high-order LB model proposed in \cite{MachadoPaper3Moments1D} 
for $z$ up to 4 and $d=1,2, 3$.  

\section{On the Prandtl number}\label{section:Prandtl}

Based on an approach to obtain the macroscopic relations, e.g. method of moments on the Boltzmann equation \eqref{eq:BoltzmannEq} and using the relations \eqref{eq:AccuracyD1Q51D} and \eqref{eq:AccuracyD2QnqCentral}, the mass continuity  and the Navier-Stokes equations, 
\eqref{eq:MassConservation}, \eqref{eq:MomentumConservation}, can be obtained (as in \cite{MachadoPaper3Moments1D}). 
The resulting equation concerning the conservation of energy is 
included in this work. That is, a total of $d+2$ relations, where $d=3$: 
\begin{subequations} \label{eq:MacroscopicRelationsConservation}
\begin{eqnarray} \label{eq:MassConservation}
\partial_t \rho + \partial_{\alpha} j_{\alpha} &=& 0, \\ \label{eq:MomentumConservation}
\rho \partial_t u_{\alpha} +  \rho u_{\beta} \partial_{\beta} u_{\alpha}  &=& -  \partial_{\alpha} p  + 
\partial_{\beta} \Bigg[ \nu \Bigg( \partial_{\beta} u_{\alpha} \\ \nonumber
&+& \partial_{\alpha} u_{\beta} - \Big( \frac{2}{3} \Big)   (\partial_{\Omega} u_{\Omega}) \delta_{\alpha, \beta} \Bigg) \Bigg], \\
\label{eq:EnergyConservation}
\frac{3 \rho}{2}\Big(\partial_t \theta^{*} + u_{\alpha} \partial_{\alpha} \theta^{*}\Big) &=&
- \rho \theta^{*} \partial_{\alpha} u_{\alpha} - \partial_{\alpha} q_{\alpha} \\ \nonumber
&-& \sigma_{\alpha, \beta} \partial_{\beta} u_{\alpha}. \
\end{eqnarray}
\end{subequations}
$p= \rho (2 \mu +1) \theta$ is the pressure, while $\nu = \rho (2 \mu +1) \theta \tau^{*}$ is the kinematic viscosity, i.e. the $\theta$ value is re-scaled by $(2 \mu +1)$. This outcome can also be valid for \eqref{eq:LB}, where $\tau^{*} = (\tau - 1/2)$, as seen in the literature, c.f. Eq. (7) in \cite{MachadoMATCOMBC}, Eq. (22) in \cite{MachadoPaper3Moments1D}. 
 The re-scaling hints $\theta^{*} = \theta (2 \mu +1)$ in \eqref{eq:EnergyConservation} and 
 with re-scaled heat flux $q_{\alpha}=-5/2 \tau^{*} p (2\mu+1)  \partial_{\alpha} \theta=-\kappa^{*} \partial_{\alpha} \theta$ and stress tensor $\sigma_{\alpha, \beta} = -2 \tau^{*} p \partial_{\alpha} u_{\beta}=-2\mu^{*}\partial_{\alpha} u_{\beta}$,  and later   dividing Eq. \eqref{eq:EnergyConservation} by $(2 \mu +1)$ leads to the Pr number (dimensionless number of the ratio between $\mu^{*}$ and $\kappa^{*}$): 
 \begin{eqnarray} \label{eq:FlexiblePrandtlNumber}
  \textrm{Pr} = (2\mu+1)^{-1}. 
\end{eqnarray}
 For instance, for the case of Pr = 2/3, the $\mu=1/4$, which is an admissible variable in the proposed construction in \cite{MachadoPaper3Moments1D}. 
 A comparison between one of the previous models, e.g. the popular ES-BGK model where the Pr=$(1-b)^{-1}$, and 
 Eq. \eqref{eq:FlexiblePrandtlNumber} shows where they coincide. 
 
 That is, both the stress tensor and heat flux are modified when the new system of moment (c.f. Eqs. \eqref{eq:AccuracyD1Q51D}) is implemented.  
 

\section{Further discussions}\label{section:Discussions}
 
Because of the novelty of the construction, obviously, there are still many open questions regarding the stability of the method in thermal mode. A theoretical study related to the H-theorem is not necessarily a trivial one. 
It deserves a separate work, which is presented by the author elsewhere. Even so, there is no guarantee of stability. 
For instance, despite the theoretical outcome in \cite{AndriesESM2000}, the ES-BGK-Burnett equations are unstable with the correct Pr number but stable when the plain BGK-Burnett equations are implemented \cite{Balakrishnan1999}, \cite{AgarwalYunBalakrishnan2001}, \cite{AgarwalYun2002}, \cite{ZhengStruchtrup2004}. 
On the other hand, rather than a construction dealing with a limited number of modified moments as in \cite{Holway63}, \cite{Shakhov68}, \cite{GLiu90}, \cite{XuHuang2010}, a new \textit{system} (i.e. Eqs. \eqref{eq:GeneralizedMoment}, \eqref{eq:AccuracyD1Q51D}, \eqref{eq:AccuracyD2QnqCentral} and (50)-(51) in \cite{MachadoPaper3Moments1D}) is adopted 
in this work. 

If the LB construction in  \cite{MachadoPaper3Moments1D} is the chosen route for the numerical implementation, it should be pointed out that:  
$i$) There exist admissible $\mu$ values so that positive populations can be theoretically obtained over several ranges of valid $\theta$ (extreme) values. Admissible lattice flow velocities at those $\theta$ extreme values can also be found. They are presented elsewhere when needed, as these ranges depend on the implemented lattice set; 
$ii$) Although the choice of finite difference schemes is usually due to stability reasons, their implementations contradict the essence of the main LB idea (c.f.  \cite{SucciBook}, \cite{BrownleeGorbanLevesley2008}); 
$iii$) As covered in the aforecited works on MRT, the use of more than one single relaxation time can enhance stability, which is always desirable. 
Because of the present construction gives already a flexible Prandtl number, the construction of a scheme with more than one relaxation time  
should focus mainly on stability. 
A work on this is also the subject of study by the author elsewhere.  

\begin{figure}[!htbp]
 \begin{center}
\includegraphics[angle=0, width=0.4\textwidth]{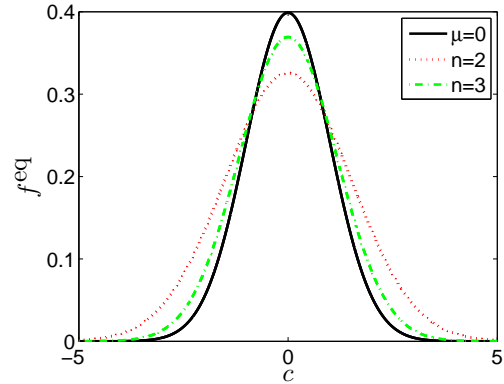}
    \caption{EDF ($f^{\textrm{eq}}$), i.e. $I$-term in Eq. \eqref{eq:GeneralizedMoment}, with $u=0$, $m=d=\rho=\theta=1$ and MB: $\mu=0$; Present: $\mu=1/4$, $n=2, 3$. 
    }
    \label{fig:ContinuousMachadoDistribution}
  \end{center}
\end{figure}

In essence, this work reaffirms the \textit{theoretical} superiority of the moment system proposed by Machado in \cite{MachadoPaper3Moments1D}, whose integral generating form is outlined here, over the MB moment system, 
at least when handling with the Eqs. \eqref{eq:MacroscopicRelationsConservation} (in both isothermal mode in \cite{MachadoPaper3Moments1D} and in thermal mode in this work).  
This is achieved by changing the targeted local equilibrium state, from the MB distribution function (c.f. $\mu=0$ in Fig. \ref{fig:ContinuousMachadoDistribution}), to 
distributions originally proposed by Machado in \cite{MachadoPaper3Moments1D} in discrete form, and  suggested in this work 
in continuous form. 

Because of the proposed construction is a system, this opens for further works beyond the Navier-Stokes-Fourier equations, which are presented by the author elsewhere.

\maketitle

\bibliography{RaulMACHADO4}%

\end{document}